\documentclass[twocolumn,prb,amsmath,amssymb,amsfonts,superscriptaddress,floatfix,nopacs]{revtex4}
\usepackage{graphicx}
\usepackage{dcolumn}
\usepackage{bm}
\usepackage{rotating}
\usepackage{color}

\def\be{\begin{equation}}
\def\ee{\end{equation}}
\def\bd{\begin{displaymath}}
\def\ed{\end{displaymath}}
\def\-{\phantom{-}}

\begin{document}

\title{First principles investigation of nanopore sequencing using variable voltage bias on graphene-based nanoribbons}

\date{\today}

\author{Hannah L. McFarland}
\affiliation{Department of Biology, James Madison University, Harrisonburg, VA 22802, USA}
\author{Towfiq Ahmed}
\affiliation
{Theoretical Division, Los Alamos National Laboratory, Los Alamos, New Mexico 87545}
\author{Jian-Xin Zhu}
\affiliation
{Theoretical Division, Los Alamos National Laboratory, Los Alamos, New Mexico 87545}
\affiliation
{Center for Integrated Nanotechnologies, Los Alamos National Laboratory, Los Alamos, New Mexico 87545}
\author{Alexander V. Balatsky}
\affiliation
{Institute of Material Science, Los Alamos National Laboratory, Los Alamos, New Mexico 87545}
\affiliation
{Nordic Institute for Theoretical Physics, KTH Royal Institute of Technology and Stockholm University, Roslagstullsbacken 23, 106 91 Stockholm, Sweden}
\author{Jason T. Haraldsen}
\affiliation{Department of Physics and Astronomy, James Madison University, Harrisonburg, VA 22802, USA}

\begin{abstract}

In this study, we examine the mechanism of nanopore-based DNA sequencing using a voltage bias across a graphene nanoribbon. Using density function theory and a non-equilibrium Green's function approach, we determine the transmission spectra and current profile for adenine, guanine, cytosine, thymine, and uracil as a function of bias voltage in an energy minimized configuration. Utilizing the transmission current, we provide a general methodology for the development of a three nanopore graphene-based device that can be used to distinguish between the various nucleobases for DNA/RNA sequencing. From our analysis, we deduce that it is possible to use different transverse currents across a multi-nanopore device to differentiate between nucleobases using various voltages of 0.5, 1.3, and 1.6 V. Overall, our goal is to improve nanopore design to further DNA/RNA nucleobase sequencing and biomolecule identification techniques. 

\end{abstract}

\maketitle

\section{Introduction}

Advances in medicine and the understanding of biomolecule interactions is on the forefront of DNA (deoxyribonucleic acid) sequencing research\cite{jone:02,ehri:02}. Errors or mutations in the nucleobase coding sequence have been shown to lead to the deregulation of gene products including proteins and enzymes\cite{grif:99book}. These coding errors can be caused by various environmental factors like radiation or chemical exposure and even viruses, which can lead to many diseases like cancer\cite{griv:10}. Therefore, there is a unique need to identify variations in genes with the resolution of individual nucleobase sequences, which can help lead to advances in medicines and therapies for a wide variety of genetic-related diseases as well as cancer\cite{smit:95}. With the human genome project working towards the ability to sequence hundreds of genomes more efficiently and cost effectively\cite{coll:03, adam:91}, the ability to sequence DNA or RNA (ribonucleic acid) down to the individual nucleobase is in great demand. Technologies that can help sequence and compare multiple genomes more accurately could lead to the possibility of diagnosing and developing treatments for patients within a �personalized medicine� approach.

There are numerous approaches for nucleobase sequencing\cite{maxa:77,stad:79,bras:03}, where the most common technique is the Sanger sequencing method that uses a chain termination approach\cite{sang:77}. More recently, there have been a number of next generation sequencing methods that have gained attention\cite{bren:00,shen:05}, but many of these methods are either inefficient and/or very expensive. However, one major avenue for individual base sequencing is the use of nanopore technology, where a single DNA/RNA strand (ssDNA or ssRNA) is drawn through a nanoscopic pore in a conductive material and measurements of variations in the optical or electronic properties can be used to identify the individual bases. 

\begin{figure}[b]
\includegraphics[width=3.25in]{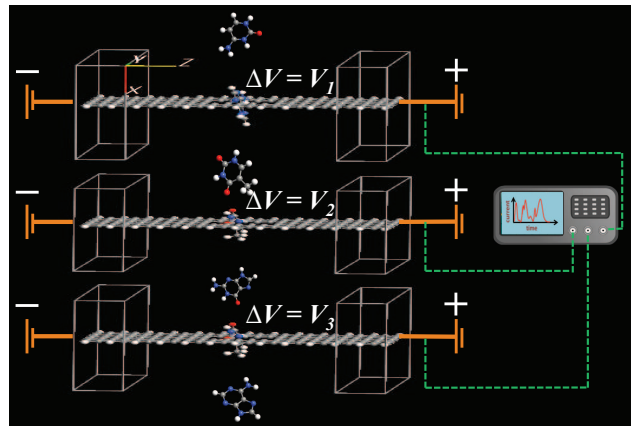}
\caption{Schematic representation of nanopore sequencing device consisting of three graphene nanoribbons.}
\label{device}
\end{figure}

Nanopore-based technology has the potential to be an efficient method for nucleobase sequencing\cite{kili:07,kili:11,tana:09,wanu:11,gara:10,shim:13}, as well as an identifier of other biomolecules for various biological sensors. There have a number of realizations of this technology using gold and other material substrates\cite{path:12}. However, the thickness of the nanopore is critical for the identification of individual nucleobases due to resulting noise and resolution problems\cite{tsut:10,chan:10,ohsh:12}, which has led to a number of publications suggesting the use of 2D materials for nanopores because the atomically-thin materials provide the proper resolution needed for individual base identification\cite{ahme:14nano,ahme:14pcl}. However, while nanopores can yield a superior resolution, there are many challenges regarding strength, durability, and overall systematic noise. Therefore, the design and material makeup of the device must address these challenges.

\begin{figure*}
\includegraphics[width=5.5in]{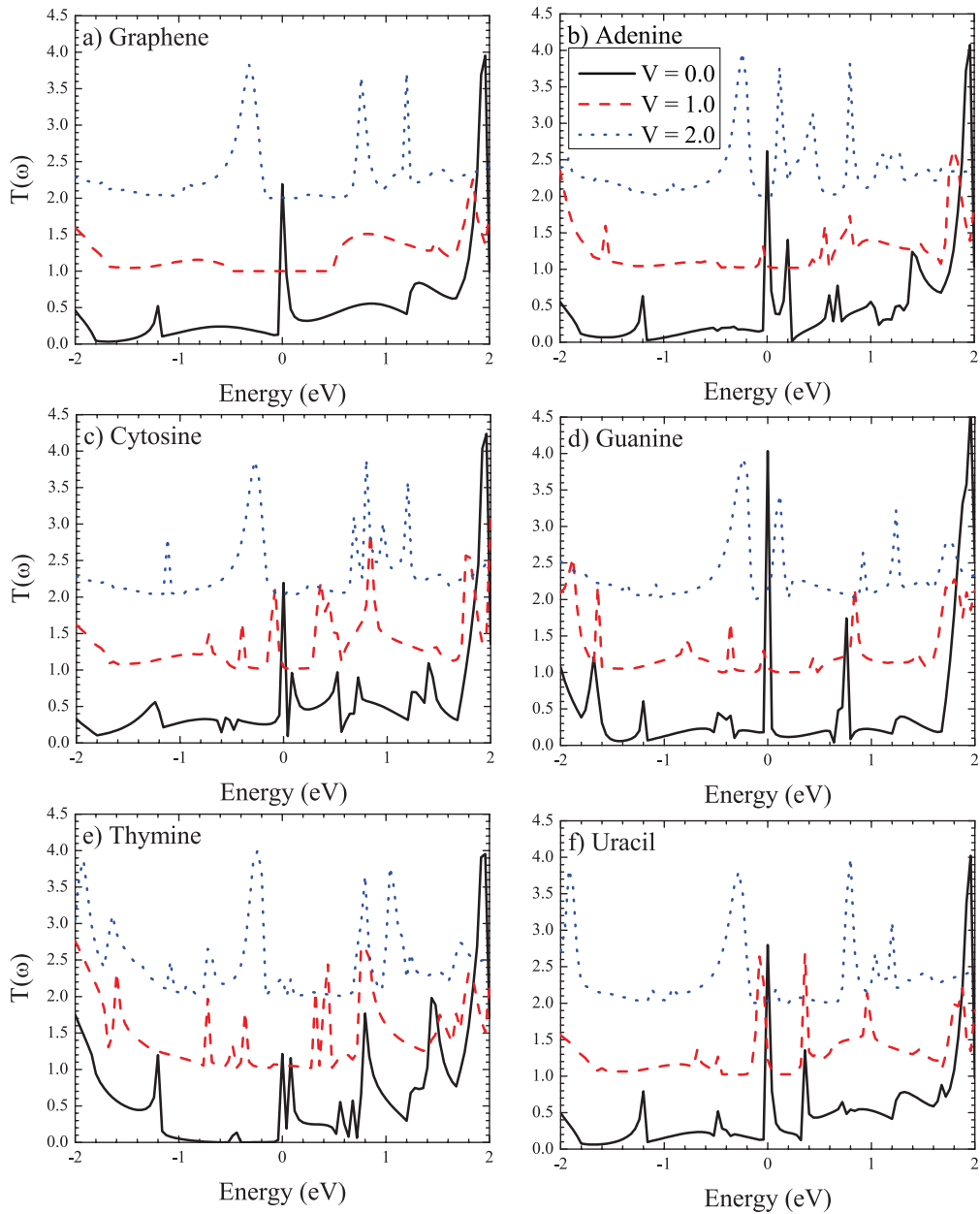}
\caption{Transmission coefficient as a function of energy for the graphene nanoribbon with an empty nanopore (a), adenine (b), cytosine (c), guanine (d), thymine (e), and uracil (f) in the nanopore.}
\label{trans}
\end{figure*}

Recently, graphene has been suggested as a possible 2D nanopore material due to its increased electron mobility and its high tensile strength\cite{ahme:14nano,ahme:14pcl}. Graphene consists of an atomically thin layer of carbon atoms arranged in a honeycomb lattice (Fig. \ref{device})\cite{geim:07}, where the lattice structure and sp$_2$ hybridized 2D graphene sheet with dangling $p$-orbitals contributes to its large tensile strength and allows for conductivity of the valence electrons\cite{cast:09,lee:08}. Therefore, the use of graphene nanopores will provide the electronic conduction and material strength needed while also having the necessary atomic resolution. While this addresses the durability and strength issues, it has been shown that the device setup can greatly reduce the general noise to signal ratio through the use of multiple sequential nanopore ribbons (shown in Fig. \ref{device}), where the use of three separate nanopores help to reduce and distinguish random and systematic noise parameters\cite{ahme:14nano}. 

In this study, we examine the viability of a multi-nanoribbon graphene nanopore device as a nucleobase sequencer. Using density functional theory and a non-equilibrium Green's function (NEGF) method, we simulate transmission spectra and calculate the ballistic currents and tunneling conductance as functions of voltage bias for adenine (A), thymine (T), cytosine (C), guanine (G), and uracil (U) translocating through the graphene nanopore. Through the analysis of I-V curves, we show that by applying various bias voltages across a graphene ribbon for the general, energy-minimized position of the translocated nucleobase, it is possible to distinguish individual bases using the resulting current. Furthermore, we provide a general mechanism that can be used for identification of the nucleobases with specific voltage biases.

\begin{figure}
\includegraphics[width=3.0in]{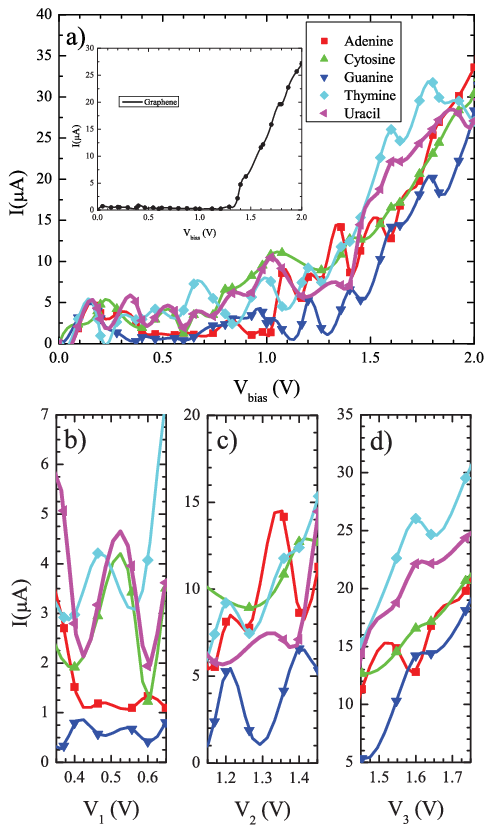}
\caption{Current as a function of voltage for all nucleobases (adenine, guanine, cytosine, thymine, and uracil) in a graphene nanopore. The inset shows the current as a function of voltage for a lone graphene nanopore ribbon. The critical voltage in the graphene is due to the presence of the nanopore and not an intrinsic property of graphene.}
\label{IV}
\end{figure}

\section{Computational Methodology}

For our simulation, we construct a graphene nanoribbon structure with a single 0.5 nm nanopore in the center. Hydrogen atoms were used to passivate the dangling bonds of the dangling orbitals of the carbon atoms at the edge of the nanoribbon and nanopore, which prevents the nucleobases from bonding to the nanoribbon and provides stability translating molecule. In addition, we place electrodes at each end of the ribbon that will simulate with a finite voltage bias (shown in Fig. \ref{device}). 

To help isolate the individual base responses, the sugar-phosphate backbone, typically found in ssDNA, is ignored, and only the individual nucleobases are placed in the nanopore. The backbone produces noise that can interrupt the current response. However, since the backbone will produce a systematic noise, previous studies have shown that analysis of cross-correlations between multiple nanoribbons can be used to reduce and possibly eliminate the systematic noise, as well as thermal and fluid fluctuations due to general flow of the DNA through the nanopore\cite{ahme:14nano}. 
%Furthermore, removing the backbone allows for the calculation to be more manageable by reducing computational time. 
Calculations were performed on a single nanopore graphene nanoribbon and one individual nucleobase in a translocating position of 60$^{\circ}$ determined through molecular dynamics simulations\cite{well:12}.

\begin{figure}
\includegraphics[width=2.75in]{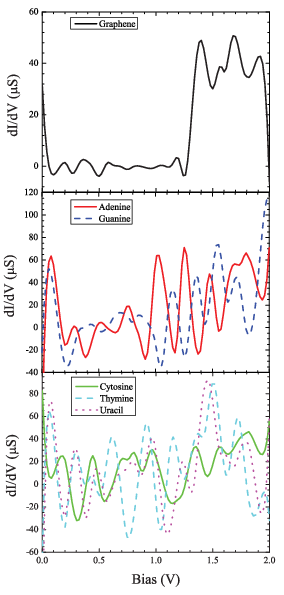}
\caption{The tunneling conductance as a function of the voltage for the nanopore and various bases translocating through the nanopore. The transmission spectrum is shown for voltage biases of 0.0 V, 1.0 V, and 2.0 V.}
\label{didv}
\end{figure}

Using Atomistix Toolkit (ATK) by Quantumwise$^{\textregistered}$, we performed density functional calculations using a generalized gradient approach (GGA)\cite{ATK,bran:02,sole:02}. From a NEGF method, we determine the transmission spectra for each base and various voltage biases. The transmission spectra is then determined from the transmission coefficient,

\be
T(\omega,V) = \sum_{{\bf k}}\sum_{nl} t_{nl}(\omega,{\bf k}, V) t_{ln}^{\dag}(\omega,{\bf k},V),
\ee

\noindent where $t_{n\ell}(\mathbf{k})$ represents the transmission amplitude from $\psi_n(\mathbf{k})$ in the left (L) electrode to $\psi_\ell(\mathbf{k})$ in the right (R) electrode. This is determined through a calculation of the Kohn-Sham Hamiltonian and the density matrix, which is given as

\be
\begin{array}{l}
\displaystyle \bar D = \frac{1}{\pi}\int_{-\infty}^{\mu_L} \bar G(\omega) {\rm Im} [\bar \Sigma^L]  \bar G(\omega)^{\dag} d\omega\\ \\
\displaystyle + \frac{1}{\pi}\int_{-\infty}^{\mu_R} \bar G(\omega) {\rm Im} [\bar \Sigma^R]  \bar G(\omega)^{\dag} d\omega.
\end{array}
\ee
\noindent Here, $\bar G(\omega)$ is the retarded Green's function and $\bar \Sigma$ is the self energy for the left and right electrodes\cite{stok:08}. The chemical potentials of the left electrode,  $\mu_L = E_F^L-e V_L$  and the right electrode, $\mu_R =E_F^L-e V_R$ are defined relative to the Fermi level of the left electrode $E_F^L$. and related to the applied bias through $\mu_R - \mu_L = e V$ and $V = V_L - V_R$.

Using various simulated voltages, we determined the current versus voltage for each nucleobase in the nanopore. The current is independent of how the left and right voltages are applied and only depends on voltage bias or difference.

\section{Results and Discussion}

Figure \ref{trans} shows the calculated transmission coefficients as a function of energy for the graphene nanoribbon with a single nanopore for V = 0.0, 1.0, and 2.0 V. The different panels show the various base configurations in the nanopore (empty, adenine, cytosine, guanine, thymine, and uracil). In this calculation of the transmission spectrum, the background contribution from the large phosphate backbone is ignored, since the background noise from the heavy and rigid backbone structure can be identified and subtracted from the general spectra.

Through an integration of transmission coefficient, the ballistic current for nucleobase can be calculated as a function of voltage

\be
I (V) = \frac{e}{h}\int_{-\infty}^{\infty}T(\omega,V)\big[ n_F (\omega - \mu_L) - n_F (\omega - \mu_R)\big] d\omega
\ee

where $n_F$ is the Fermi function and voltage is defined as the difference between the left and right electrode chemical potentials,  $\mu_R - \mu_L = e V$.  

In Fig. \ref{IV}(a), we show the calculated current as a function of voltage for the various nucleobases (adenine, thymine, guanine, cytosine, and uracil), where the empty graphene nanopore is presented in the inset. From this data, there are distinct voltage pathways for the differentiation of all bases. The sudden rise in current at 1.3 V is due to the presence of the nanopore in the simulation and can be traced back to the graphene itself. Since the simulation assumes a nanopore on the size order of the nanoribbon itself, there is a critical voltage for which the empty graphene nanopore will not produce a current due to a structurally induce energy gap. Therefore, once 1.0 V is achieved, electrons can overcome this energy barrier and produce a sizable current. If no nanopore existed, then the graphene nanoribbon would have its normal conductivity. Therefore, the presence of a nucleobase allows for current to be drawn through the nanoribbon at voltage differences lower than 1.0 V.

From the current, the tunneling conductance (dI/dV) (shown in Fig. \ref{didv}) can be determined through a differentiation of the I-V curve. Therefore, there is the possibility that topological probes can be used to examine the individual bases as well. This may be useful for scanning tunneling microscopy. 

\begin{table}
\caption{General response for each DNA and RNA nucleobase in a three nanopore setup.}
\begin{tabular}{lccc}
\hline
Nanobase & $V_{1}$ (0.5 V) & $V_{2}$ (1.3 V) & $V_{3}$ (1.6 V) \\
\hline
Adenine & low & high & low \\
Guanine & low & low & low \\
Cytosine & high & high & low \\
Thymine (DNA) & high & high & high \\
Uracil (RNA) & high & high & high \\
\hline
\end{tabular}
\label{tab1}
\end{table}

From Fig. \ref{IV}(a), there are specific voltages that produce distinct current variations. Therefore, in order to produce a nanopore device that can distinguish individual nucleobases, the use of multiple nanoribbons is needed. This is not a problem since multiple nanoribbons are needed for the appropriate noise reduction. Figures \ref{IV}(b)-(d) zoom in around the characteristic voltages (V$_1$, V$_2$, V$_3$ as illustrated in Fig. \ref{device}) for nucleobase differentiation are 0.5, 1.3, and 1.6 V, respectively. Through a comparison to background, the utilization of these voltages allows us to distinguish each base by evaluating a high or low signal or current response. This is shown in Table \ref{tab1} for all nucleobases.

\begin{figure}
\includegraphics[width=3.25in]{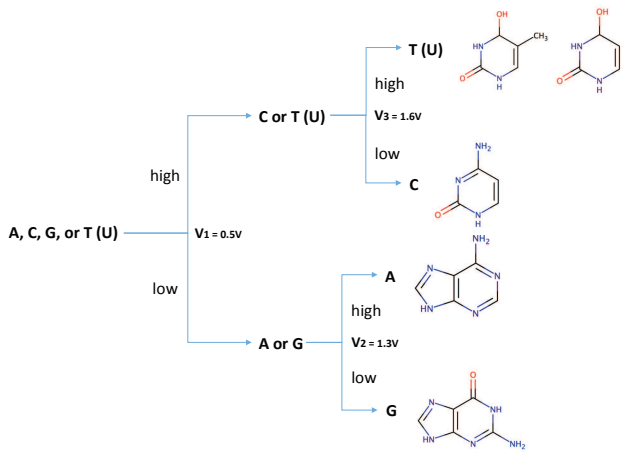}
\caption{Illustrates the differentiation pathway for the bases translocating through three nanopores for DNA bases (a) and RNA bases (b). This reveals the distinct possibility for the use of graphene nanopores as a sequencing device.}
\label{scheme}
\end{figure}

Figure \ref{scheme} illustrates the methodology needed to characterize the bases. The first nanoribbon will provide a small voltage bias (V$_1$ = 0.5 V), and will be able to determine pairs of bases by have a low (A or G) or high (C or T) current compared to the normalized baseline. The nucleobase will then pass through a second nanoribbon with a moderate voltage bias (V$_2$ = 1.3 V), which can be used to identify either G (low) or A (high) assuming a low first voltage. The third nanoribbon at a higher voltage bias (V$_3$ = 1.6 V) will be used to determine C (low) and T (or U) (high). Once the nucleobase has translocated through all three nanopores, the characteristic current sequence will allow for the identification of the individual base.

\section{Conclusions}

Graphene has been shown to be a potential material for nanopore-based sequencing, due to its atomic thickness and relative strength. Using density functional calculations, we find that the five nucleobases, including uracil, can be distinguished through the use of multiple nanoribbons using variable voltage biases. From the simulated transmission spectra, we calculate the I-V curves for these nucleobases. By examining specific differences in the calculated current, the precise nucleobase that is translocating through the nanopore can be determined. We focus on voltages of 0.5, 1.3, and 1.6 V as a proof of principle for a specific nanopore sequencing device. 

Future work includes performing conductance calculations for specific voltages for a better microscopic understanding as well as looking at surface plasmon resonances. In addition, increasing the size of the calculation, such as a large nanopore or a full DNA strands calculation. Further investigations include a time-dependent translocation through the nanopore that includes thermal fluctuations. Here, we focused on graphene as possible 2D materials. However, there should also be a push forward with other 2D materials for comparison.

\section*{Acknowledgement}

H.L.M. and J.T.H. thank the support of James Madison University and useful discussions with J.-H. Kim. The work at Los Alamos National Laboratory was carried out under the auspice of the U.S. DOE and NNSA under Contract No. DEAC52-06NA25396 and supported by U.S. DOE Basic Energy Sciences Office (T.A. and A.V.B.). This work was also, in part, supported by the Center for Integrated Nanotechnologies, a U.S. DOE Office of Basic Energy Sciences user facility (J.-X.Z).

\end{document}